\documentclass[epj]{svjour}
\usepackage{graphics}
\usepackage{amsmath}
\usepackage{amssymb}
\RequirePackage{graphicx}

\usepackage[english]{babel}
\usepackage[latin1]{inputenc}
\usepackage{amsmath,amssymb}
\usepackage{bm,graphics,color}

\begin{document}

\title{A Conducting surface in Lee-Wick electrodynamics}

\author{F.A. Barone\inst{1}\thanks{e-mail: fbarone@unifei.edu.br}
        \and
        A.A Nogueira\inst{2}\thanks{e-mail: nogueira@ift.unesp.br}
					}

\institute{IFQ - Universidade Federal de Itajub\'a, Av. BPS 1303, Pinheirinho, Caixa Postal 50, 37500-903, Itajub\'a, MG, Brazil.
           \and
           IFT - Rua Dr. Bento Teobaldo Ferraz, 271, Bairro: Barra-Funda, 01140-070, S\~ao Paulo, SP, Brazil.
          }

\date{Received: date / Accepted: date}

\abstract{
The Lee-Wick electrodynamics in the vicinity of a conducting plate is investigated. The propagator for the gauge field is calculated and the interaction between the plate and a point-like electric charge is computed. The boundary condition imposed on the vector field is taken to be the one that vanishes, on the plate, the normal component of the dual field strength to the plate. It is shown that the image method is not valid in Lee-Wick electrodynamics.
%
} 

\maketitle

\section{Introduction}

The simplest higher-order derivative gauge theory is the so called Lee-Wick electrodynamics \cite{Podolsky42,Podolsky44,Podolsky48,LW69,LW70,CusMElPomp}, which is described by the Maxwell Lagrangian augmented by a higher-order derivative kinetic term. Since its proposal, the theory has been standing out by its classical as well as its quantum aspects, as it exhibits many interesting peculiarities. We can mention, for instance, the fact that in this electrodynamics the self-energy of a point-charge is finite in $3+1$ dimensions \cite{BHN2013,KruglovJPA2010,Frankel,Zayats,SantosMPLA2011,AcciolyPRD2004}, a Dirac string can produce a magnetic field \cite{BHN2013}, it stem a finite theory closely related to the Pauli-Villars regularization scheme  \cite{LW70,KruglovJPA2010,StellePRD77,Grinstein2008,GrinsteinPRD2008,KraussPRD2008,Rodigast2009,AcciolyMPLA2011} where the divergences of the quantum electrodynamics are controllable \cite{Buf1,Buf2,Buf3} and it exhibits classical dynamical stability \cite{Russos}. These features have ma\-de the model a widely studied subject in a variety of scenarios, mainly regarding its extension to the Standard Model. \cite{Grinstein2008,KraussPRD2008,AcciolyMPLA2011,Espinosa2008,Underwood2009,CaronePLB2008,CaronePLB2009,CaronePRD2009,CaroneJHEP2009,RizzoJHEP2008,RizzoJHEP2007,Schat2008,GOWPRD2008,GabrielliPRD2008,CuzinattoIJMPA2011,AcciolyMPLA2010,Accioly2010,Roman,Shalaby2009}.

In spite of all this interest, as far as the authors know, there is a lack in the literature regarding the Lee-Wick electrodynamics under the influence of boundary conditions. That is a remarkable subject in any abelian gauge theory, since the experimental apparatus commonly used to test electromagnetic phenomena are, usually, surroun\-ded by conductors. In addition, it is also important for the investigation of situations where it is possible to find deviations from expected physical results in comparison to Maxwell electrodynamics. The presence of conductors can create suitable scenarios for this kind of search.

In the work of reference \cite{Ucraniano} the presence of conducting surfaces was investigated in the context of the Casimir effect for Lee-Wick electrodynamics. In the work of reference \cite{Ravndal}, among other terms, a Lee-Wick type contribution (called Uehling term) was inserted to compose an effective theory in order to calculate radiative corrections to the Casimir effect. In this paper we highlight that the real role of a conductor in this theory is not such a common issue and requires more cautious attention. 

As is well known, this electrodynamics exhibits two modes, one massive and the other one massless. The propagator can then be splitted up into the sum of two parts, the first one being just the usual Maxwell propagator and the second one the Proca propagator with an overall minus sign. Despite the presence of these ghosts modes, as is the case of Pauli-Villars regulators, the theory can be rendered unitary provided that the Lee-Wick particles decay. This fact means that the physical effects at tree level are trivial in most cases. 

In this paper we show that this is not the case when the presence of a conducting plate is taken into account. The correction term that has to be added to the propagator in this case is not a simple subtraction of the corresponding Maxwell and Proca propagators coupled to the conductor. Due to this fact, we show that the physical phenomena are no longer as trivial as in the theory without the conducting surfaces. 

Specifically, in Section (\ref{secpropagador}) we compute the propagator for the Lee-Wick gauge field in the presence of a conducting plane. We employ Quantum Field Theory methods in order to obtain the functional generator for the gauge sector, since with this functional one can find any physical quantity of interest of the theory. In section (\ref{secespelhocarga}) we calculate the interaction between a perfectly conducting plate and a point-like charge. We also compare the results with the standard Maxwell electrodynamics and show that in the Lee-Wick model the image method is no longer valid. In section (\ref{2campos}) we make a discussion regarding the two field formalism of the Lee-Wick Theory in the presence of a conducting plate. That gives an alternative way to understand why the image method is not valid in the Lee-Wick Theory. Section (\ref{conclusoes}) is devoted to our final comments.

\section{Lee-Wick propagator in the presence of a conducting plate}
\label{secpropagador}

In this paper we work in a $3+1$ dimensional Minkowski spacetime with metric $(+,-,-,-)$. The vector gauge field is designated by $A^{\mu}$ and its corresponding field strength by $F^{\mu\nu}=\partial^{\mu}A^{\nu}-\partial^{\nu}A^{\mu}$.

The electromagnetic sector of the so called Lee-Wick electrodynamics is described by the Lagrangian density \cite{Podolsky42,Podolsky44,Podolsky48,LW69,LW70,BHN2013}	
\begin{equation}
\label{defL}
{\cal L}_{LW}=-\frac{1}{4}F_{\mu\nu}F^{\mu\nu}-\frac{1}{4m^{2}}F_{\mu\nu}\partial_{\alpha}{\partial}^{\alpha}F^{\mu\nu}-\frac{{(\partial_{\mu}A^{\mu})}^2}{2\xi}-J_{\mu}A^{\mu}\ ,
\end{equation}
where $m$ is a parameter that has mass dimension, $J^{\mu}$ is an external source and $\xi$ is a gauge fixing parameter. It is important to mention that there are other covariant gauge conditions for this theory \cite{BMpimentel,Nonmix}.

As discussed in many works \cite{Podolsky42,Podolsky44,Podolsky48,LW69,LW70,BHN2013}, the Lagran\-gian (\ref{defL}) exhibits gauge invariance and two distinct poles, in momenta space, for the corresponding propagator; a massless one and a massive one. This fact can be shown by considering that, in the Feynman gauge where $\xi=1$, the model (\ref{defL}) is equivalent to
\begin{equation}
\label{Lalternativo}
{\cal L}_{LW}\to\frac{1}{2}A_{\mu}\Bigl[\eta^{\mu\nu}\Bigl(1+\frac{\partial^{\gamma}\partial_{\gamma}}{m^{2}}\Bigr)\partial^{\alpha}\partial_{\alpha}-\frac{\partial^{\gamma}\partial_{\gamma}}{m^{2}}\partial^{\mu}\partial^{\nu}\Bigr]A_{\nu}\ ,
\end{equation}
so that the corresponding Feynman propagator, is given by
\begin{eqnarray}
\label{propagador}
D_{\mu\nu}(x,y)=\int \frac{{d^{4}p}}{(2\pi)^{4}}\Biggl(\frac{1}{p^{2}-m^{2}}-\frac{1}{p^{2}}\Biggr)\cr\cr
\biggl({\eta}_{\mu\nu}-\frac{p_{\mu}p_{\nu}}{m^{2}}\biggr)e^{-ip(x-y)}
\end{eqnarray}
in the sense that
\begin{eqnarray}
\label{propagador2}
\Bigl[\eta^{\mu\nu}\Bigl(1+\frac{\partial^{\gamma}\partial_{\gamma}}{m^{2}}\Bigr)\partial^{\alpha}\partial_{\alpha}-\frac{\partial^{\alpha}\partial_{\alpha}}{m^{2}}\partial^{\mu}\partial^{\nu}\Bigr]D_{\nu\beta}(x,y)=\cr\cr
=\eta^{\mu}_{\ \beta}\delta^{4}(x-y)\ ,
\end{eqnarray}
where it is implicit, from now on, that there is a small imaginary part for the momentum square, $p^{2}\to p^{2}+i\varepsilon$.

The Maxwell and Lee-Wick electrodynamics yield different dynamical equations for the gauge field, but the dynamical equations for the charged particles and the Lorentz force are the same in both theories, namely, $\frac{d{\bf p}_{p}}{dt}=q{\bf E}+q{\bf v}\times{\bf B}$, where $q$ and ${\bf p}_{p}$ are, respectively, the charge and the spatial component of the particle momentum.

In this section we consider general aspects of the Lee-Wick electrodynamics in the presence of a conducting plate. That is a non-trivial task since its inception, because we have to establish what is a conductor in this model. 
To answer this question we resort to the behavior of the electromagnetic field in the presence of a conductor, according to the Maxwell theory.

A conducting surface in the Maxwell electrodynamics imposes a boundary condition on the gauge field in such a way the Lorentz force on the surface vanishes. It is achieved by taking as zero the component of the dual field strength, normal to the surface. That is, if $n^{\mu}$ is the normal four-vector to the conducting surface, we must have $n^{\mu}{F^{*}}_{\mu\nu}=0$ along it, where ${F^{*}}^{\mu\nu}=(1/2)\epsilon^{\mu\nu\alpha\beta}F_{\alpha\beta}$ is the dual to the field strength, with $\epsilon^{\mu\nu\alpha\beta}$ standing for the Levi-Civita tensor ($\epsilon^{0123}=1$). Once in Lee-Wick electrodynamics the Lorentz force is the same one as in Maxwell theory, the condition which vanishes the Lorentz force in both theories must be the exactly same.

Here we consider the presence of a single perfectly conducting plate. With no loss of generality we take a coordinate system where the plate is perpendicular to the $x^{3}$ axis, lying on the plane $x^{3}=a$, so its normal four-vector is $n^{\mu}=\eta_{3}^{\ \mu}=(0,0,0,1)$ and the boundary condition for the gauge field $A^{\mu}$ reads     
\begin{equation}
\label{condcondutor}
{F^{*}}_{3\nu}(x)|_{x^{3}=a}=0
\end{equation}
where the sub-index indicates that the boundary conditions are taken on the plane $x^{3}=a$.

We shall compute the functional generator for the vector field submitted to the boundary conditions (\ref{condcondutor}) following the procedure proposed in reference \cite{Bordag}, that is, carrying out the functional integral
\begin{equation}
Z_{C}[J_{\mu}]=\int{\cal D}A_{C}\ e^{i\int d^{4}x\ ({\cal L}_{LW}-J^{\mu}A_{\mu}})
\end{equation}
where the sub index $C$ means that the integral is restricted only to field configurations which satisfy the conditions (\ref{condcondutor}). This restriction is achieved with the insertion of a delta functional that is non-vanishing only for field configurations that satisfy the conditions (\ref{condcondutor}) and integrating in all field configurations, that is, 
\begin{equation}
\label{fgccc}
Z_{C}[J_{\mu}]=\int{\cal D}A\delta[{F^{*}}_{3\nu}(x)|_{x^{3}=a}]\ e^{i\int d^{4}x\ ({\cal L}_{LW}-J^{\mu}A_{\mu})}\ .
\end{equation}

Now we use the functional Fourier representation
\begin{eqnarray}
\label{deltafuncional}
\delta[{F^{*}}_{3\nu}(x)|_{x^{3}=a}]=\cr\cr
\int{\cal D}{B}\exp\Biggl[i\int d^{4}x \delta(x^{3}-a)B_{\nu}(x_{\parallel}){F^*}_{3}^{\ \nu}(x)\Biggr]
\end{eqnarray}
where $B_{\nu}(x_{\parallel})$ is an auxiliary vector field and the notation $x_{\parallel}=(x^{0},x^{1},x^{2})$ is used for the coordinates parallel to the plate.

The auxiliary field $B_{\nu}(x)$ exhibit gauge symmetry
\begin{equation}
\label{calibreB}
B_{\nu}(x_{\|})\to B_{\nu}(x_{\|})+\partial_{\nu\|}\Lambda^{(k)}(x_{\|})\ ,
\end{equation}
what requires a cautious treatment of the integral (\ref{deltafuncional}). In the appendix we show that
\begin{eqnarray}
\label{deltafuncional2}
\delta[{F^{*}}_{3\nu}(x)|_{x^{3}=a}]=\cr\cr
N\int{\cal D}B\exp\Biggl(-i\int d^{4}x\delta(x^{3}-a)A_{\beta}(x)\epsilon_{3}^{\ \nu\alpha\beta}\partial_{\alpha}B_{\nu}(x)\Biggr)\cr\cr
\exp\Biggl(\frac{i}{2\gamma}\int d^{4}xd^{4}y \delta(x^{3}-a)B^{\mu}(x_{\parallel})\cr\cr
\times\frac{\partial^{2}Q(x,y)}{\partial x_{\parallel}^{\mu}\partial y_{\parallel}^{\nu}}\delta(y^{3}-a)B^{\nu}(y_{\parallel})\Biggr)\ ,\ \ \ \ \  
\end{eqnarray}
where $\gamma$ is a gauge fixing term, $N$ is a constant which does not depend on the fields and $Q(x,y)$ is an arbitrary function that shall be chosen conveniently.

Substituting (\ref{deltafuncional2}) in (\ref{fgccc}) we get
\begin{eqnarray}
\label{asd1}
Z_{C}[J]=N\int{\cal D}A{\cal D}B\ \ e^{i\int d^{4}x\ ({\cal L}_{LW}-J^{\mu}A_{\mu})}\cr
\exp{\Biggl(-i\int d^{4}x \delta(x^{3}-a)A_{\beta}(x)\epsilon_{3}^{\ \nu\alpha\beta}\partial_{\alpha}B_{\nu}(x)\Biggr)}\cr
\exp\Biggl(\frac{i}{2\gamma}\int d^{4}xd^{4}y\ \delta(x^{3}-a)B^{\mu}(x_{\parallel})\cr
\times\frac{\partial^{2}Q(x,y)}{\partial x_{\parallel}^{\mu}\partial y_{\parallel}^{\nu}}\ \delta(y^{3}-a)B^{\nu}(y_{\parallel})\Biggr)\ .\ \ \ \ \ 
\end{eqnarray}
In the first exponential we have only the $A^{\mu}$ field and in the third one, only the presence of $B^{\mu}$. The second exponential contains a coupling between $A$ and $B$. 

In order to decouple the fields $A$ and $B$ we perform the translation
\begin{eqnarray}
A^{\beta}(x)\to A^{\beta}(x)\cr
+\int d^{4}y D^{\beta}_{\ \alpha}(x,y)\delta(x^{3}-a)\epsilon_{3}^{\ \nu\gamma\alpha}\partial_{\gamma}B_{\nu}(x)\ ,\   
\end{eqnarray}
what brings the integral (\ref{asd1}) to the form
\begin{eqnarray}
\label{Zcseparado}
Z_{C}[J]=NZ_{LW}[J]\bar{Z}[J]
\end{eqnarray}
where $Z_{LW}[J]$ is the standard Lee-Wick functional generator
\begin{eqnarray}
\label{defZLW}
Z_{LW}[J]=\int{\cal D}A\ e^{i\int d^{4}x\ ({\cal L}_{LW}-J^{\mu}A_{\mu})}\cr
=Z_{LW}[0]\exp{\Biggl[-\frac{i}{2}\int d^{4}xd^{4}y J^{\mu}(x)D_{\mu\nu}(x,y)J^{\nu}(y)\Biggr]}\ ,\cr
\ 
\end{eqnarray}
which can be calculated using standard methods of quantum field theory \cite{BHN2013}, and $\bar{Z}[J]$ is a contribution that does not involve $A^{\mu}$,
\begin{eqnarray}
\label{defbarZ}
\bar{Z}[J]=\int{\cal D}B\ \exp{\Biggl[i\int d^{4}x\delta(x^{3}-a)I^{\nu}(x)B_{\nu}(x_{\|})\Biggr]}\cr\cr
\exp\Bigg[i\int d^{4}xd^{4}y\delta(x^{3}-a)\delta(y^{3}-a)B_{\nu}(x_{\|})B_{\pi}(y_{\|})\cr\cr
\Bigg(\frac{1}{2}\epsilon_{3}^{\ \nu\alpha\lambda}\epsilon_{3}^{\ \pi\gamma\rho}\Big(\partial_{\alpha}\partial_{\gamma}D_{\lambda\rho}(x,y)\Big)+\frac{1}{2\gamma}\frac{\partial^{2}Q(x,y)}{\partial x_{\|\nu}\partial y_{\|\pi}}\Bigg)\Bigg]
\end{eqnarray}
where we defined
\begin{equation}
\label{defI}
I^{\nu}(x)=-\int d^{4}y \epsilon_{3}^{\ \nu\alpha\rho}\Bigg(\frac{\partial}{\partial x^{\alpha}}D_{\rho\lambda}(x,y)\Bigg)J^{\lambda}(y)
\end{equation}

Notice that the integral (\ref{defbarZ}) is Gaussian, so that, it can be calculated exactly. For this task it is convenient to make the following choice
\begin{equation}
\label{defQ}
Q(x,y)=\int\frac{d^{4}p}{(2\pi)^{4}}\Biggl(\frac{1}{p^{2}-m^{2}}-\frac{1}{p^{2}}\Biggr)e^{-ik(x-y)}\ ,
\end{equation}
and work in the gauge where $\gamma=1$. Substituting (\ref{defQ}) and (\ref{defI}) into (\ref{defbarZ}), using the fact that 
\begin{eqnarray}
\label{asd4}
\int\frac{dp^{3}}{2\pi}\frac{1}{p^{2}-m^{2}}e^{ip^{3}x^{3}}=-\frac{i}{2\Gamma}e^{i\Gamma |x^{3}|}\cr\cr
\int\frac{dp^{3}}{2\pi}\frac{1}{p^{2}}e^{ip^{3}x^{3}}=-\frac{i}{2L}e^{iL|x^{3}| a}
\end{eqnarray}
where $\Gamma=\sqrt{p_{\|}^{2}-m^{2}}$ and $L=\sqrt{p_{\|}^{2}}$ (see appendix), defining the parallel momentum to the plate $p_{\|}=(p^{0},p^{1},p^{2},0)$ and the parallel metric
\begin{equation}
\label{defetaparalela}
\eta^{\mu\nu}_{\|}=\eta^{\mu\nu}-\eta^{\mu}_{\ 3}\eta^{\nu}_{\ 3}
\end{equation}
and carrying out some manipulations, one can write Eq. (\ref{defbarZ}) in the form
\begin{equation}
\label{barZcalculado}
{\bar Z}[J]={\bar Z}[0]\exp\Bigg[-\frac{i}{2}\int d^{4}x d^{4}y J^{\mu}(x){\bar D}_{\mu\nu}(x,y)J^{\nu}(y)\Bigg]
\end{equation}
where we defined the function,
\begin{eqnarray}
\label{defbarD}
{\bar D}^{\mu\nu}(x,y)=\int\frac{d^{3}p_{\|}}{(2\pi)^{3}}-\frac{i}{2}\Biggl(\eta_{\|}^{\mu\nu}-\frac{p_{\|}^{\mu}p_{\|}^{\nu}}{p_{\|}^{2}}\Biggr)\frac{1}{\frac{1}{L}-\frac{1}{\Gamma}}\cr
\exp[-ip_{\|}(x_{\|}-y_{\|})]\Biggl(\frac{1}{L}e^{iL|x^{3}-a|}-\frac{1}{\Gamma}e^{i\Gamma|x^{3}-a|}\Biggr)\cr
\Biggl(\frac{1}{L}e^{iL|y^{3}-a|}-\frac{1}{\Gamma}e^{i\Gamma|y^{3}-a|}\Biggr) .
\end{eqnarray}

Substituting (\ref{defZLW}) and (\ref{barZcalculado}) in (\ref{Zcseparado}) we have the functional generator of the Lee-Wick gauge field in the presence of a conducting plate
\begin{eqnarray}
\label{Zfinal}
Z[J]_{C}=Z[0]\exp\Bigg[-\frac{i}{2}\int d^{4}x d^{4}y J^{\mu}(x)\Bigl(D_{\mu\nu}(x,y)\cr
+{\bar D}_{\mu\nu}(x,y)\Bigr)J^{\nu}(y)\Bigg]\ .
\end{eqnarray}

Notice that, from the above expression (\ref{Zfinal}), one can identify the propagator of the theory in the presence of a conducting plate as,
\begin{equation}
\label{defDC}
D_{C}^{\mu\nu}(x,y)=D_{\mu\nu}(x,y)+{\bar D}_{\mu\nu}(x,y) \ .
\end{equation}

As a matter of checking we point out that the gauge field propagator under the boundary conditions (\ref{defDC}) is really a Green function for the problem, in the sense that it is the inverse of the Lee-Wick operator, obtained from Eq. (\ref{Lalternativo}), that is
\begin{eqnarray}
\Bigl[\eta^{\mu\nu}\Bigl(1+\frac{\partial^{\gamma}\partial_{\gamma}}{m^{2}}\Bigr)\partial^{\alpha}\partial_{\alpha}-\frac{\partial^{\gamma}\partial_{\gamma}}{m^{2}}\partial^{\mu}\partial^{\nu}\Bigr]D_{C}^{\mu\nu}(x,y)\cr
=\delta^{4}(x-y)\ .
\end{eqnarray}
The above equation can be verified directly by using Eq's (\ref{defDC}), (\ref{propagador2}) and (\ref{defbarD}).

Moreover, any field configuration obtained from (\ref{defDC}) satisfies the boundary condition (\ref{condcondutor}). It can be checked by calculating the field  generated by an arbitrary source
\begin{eqnarray}
\label{jkl1}
A^{\mu}(x)=\int d^{4}y D_{C}^{\mu\nu}(x,y)J_{\nu}(y)\ .
\end{eqnarray}
In terms of the gauge field $A$, the boundary conditions (\ref{condcondutor}) reads $\epsilon^{3\nu\alpha\beta}\partial_{\alpha}A^{\mu}(x)|_{x^{3}=0}=0$, so, by using (\ref{jkl1}) it can be shown that the propagator must satisfy 
\begin{equation}
\label{jkl2}
\epsilon^{3\nu\alpha\beta}\frac{\partial D_{C}^{\mu\nu}(x,y)}{\partial x^{\alpha}}|_{x^{3}=0}=0\ .
\end{equation}
With the aid of Eq's (\ref{defDC}), (\ref{propagador2}) and (\ref{defbarD}) it can be shown that the condition (\ref{jkl2}) is really satisfied.

At this point some comments are in order. The propagator (\ref{defDC}) is composed by the sum of the free Lee-Wick propagator (\ref{propagador}) with the correction (\ref{defbarD}) which accounts for the presence of the conducting plate. As exposed in the appendix, in the limit when $m=\infty$ the propagator (\ref{defDC}) reduces to the same one as that found with Maxwell electrodynamics in the presence of a conducting plate.

The free Lee-Wick propagator (\ref{propagador}) is made up by the Maxwell propagator minus the Proca one. This fact makes most results of theory to be the composition of the corresponding ones obtained in the Maxwell and in the Proca theory. When the boundary condition (\ref{condcondutor}) is involved this is no longer valid and the propagator (\ref{defDC}) is not compound by the corresponding ones of the Maxwell and Proca theories with boundary conditions. This result suggests that the boundary conditions mix up the modes of the Lee-Wick field with and without mass (the photon and its Lee-Wick partner). Due to this fact, some physical phenomena of the Lee-Wick electrodynamics in the vicinity of a conducting plate are not trivial and deserve investigation.

\section{Particle-Plate interaction}
\label{secespelhocarga}

In this section we consider the interaction between a point-like charge and a conducting plane. By using the arguments discussed in references \cite{BHN2013,GBB,BB,Anderson}, we can show that the interaction energy between a conducting surface and an external source $J^{\mu}(x)$ for the gauge field is given by the integral
\begin{equation}
\label{xcv1}
E=\frac{1}{2T}\int d^{4}xd^{4}y J_{\mu}(x)\bar{D}^{\mu\nu}(x,y)J_{\nu}(y)
\end{equation}

In our case we take the source corresponding to a point-like steady charge placed at position ${\bf b}$.
\begin{equation}
\label{defJpontual}
J^{\mu}(x)=q\eta^{\mu0}\delta^{3}({\bf x}-{\bf b})\ .
\end{equation}

Substituting (\ref{defJpontual}) in (\ref{xcv1}), using the Fourier representation (\ref{defDC}), carrying out the integrals in $d^{3}{\bf x}$, $d^{3}{\bf y}$, $dx^{0}$, $dk^{0}$ and $dy^{0}$ and making some simple manipulations we obtain,
\begin{eqnarray}
\label{xcv2}
E_{PC}=-\frac{q^{2}}{4}\int\frac{d^{2}{\bf p}_{\|}}{(2\pi)^{2}}\frac{\sqrt{{\bf p}_{\|}^{2}+m^{2}}\sqrt{{\bf p}_{\|}^{2}}}{\sqrt{{\bf p}_{\|}^{2}+m^{2}}-\sqrt{{\bf p}_{\|}^{2}}}\cr
\Biggl(\frac{\exp{(-\sqrt{{\bf p}_{\|}^{2}}\ R)}}{\sqrt{{\bf p}_{\|}^{2}}}-
\frac{\exp{(-\sqrt{{\bf p}_{\|}^{2}+m^2}\ R)}}{\sqrt{{\bf p}_{\|}^{2}+m^2}}\Biggr)^{2}\ ,
\end{eqnarray}
where we defined $R=|{\bf a}-{\bf b}|$ which stands for the distance between the plate and the charge. The sub-index $PC$ means the interaction energy between the plate and a charge.

Expression (\ref{xcv2}) can be simplified by using polar coordinates, integrating out in the solid angle and performing the change of integration variable $p=|{\bf p}_{\|}|/m$,
\begin{eqnarray}
\label{xcv3}
E_{PC}=-\frac{q^{2}}{2^{3}\pi}m\int_{0}^{\infty}dp\ p^{2}[(p^{2}+1)+p(p^{2}+1)^{1/2}]\cr\cr
\Biggl(\frac{e^{-2pmR}}{p^{2}}-2\frac{e^{-pmR}e^{-\sqrt{p^{2}+1}mR}}{p(p^{2}+1)^{1/2}}+\frac{e^{-2\sqrt{p^{2}+1}mR}}{p^{2}+1}\Biggr)\ .\ 
\end{eqnarray}

Each contribution in the integral (\ref{xcv3}) can be calculated exactly. For the first contribution we have
\begin{eqnarray}
\label{cont1}
\int_{0}^{\infty}dp\ [(p^{2}+1)+p(p^{2}+1)^{1/2}]\exp{(-2pmR)}=\cr\cr
=\frac{1+2(mR)^{2}}{4(mR)^{3}}+\frac{\pi}{4mR}[Y_{0}(2mR)-SH_{0}(2mR)]\cr\cr
+\frac{\pi}{4(mR)^{2}}[SH_{1}(2mR)-Y_{1}(2mR)]\ , 
\end{eqnarray}
where $Y$ and $SH$ stand for the Bessel function of second kind and the Struve function, respectively. 

For the third contribution to (\ref{xcv3}) we carry out the change in the integration variable $u=\sqrt{p^{2}+1}$, as follows
\begin{eqnarray}
\label{cont2}
\int_{0}^{\infty}dp\ p^2[1+p(p^{2}+1)^{-1/2}]\exp{(-2\sqrt{p^{2}+1}mR)}\cr
=\int_{1}^{\infty}\ du[u(u^{2}-1)^{1/2}+u^{2}-1]\exp{(-2umR)}\cr
=\frac{K_{0}(2mR)}{2mR}+\frac{K_{1}(2mR)}{2(mR)^{2}}
+\frac{\exp{(-2mR)}}{4(mR)^{3}}(1+2mR)\ ,\cr
\ 
\end{eqnarray}
where $K$ stands for the Bessel.

The second contribution to (\ref{xcv3}) is calculated with the change of variable $u=p+(p^{2}+1)^{1/2}$,
\begin{eqnarray}
\label{cont3}
\int_{0}^{\infty}dp\!\!& &\!\!-2p[(p^{2}+1)^{1/2}+p]\exp{[-(p+\sqrt{p^{2}+1})mR]}=\cr
&=&-2\int_{1}^{\infty}du\frac{u^{4}-1}{4u^{2}}\exp{(-umR)}\cr
&=&\exp{(-mR)}\Biggl(\frac{1}{2}-\frac{1}{2mR}-\frac{1}{(mR)^{2}}-\frac{1}{(mR)^{3}}\Biggr)\cr
& &-\frac{1}{2}(mR)Ei(1,mR)
\end{eqnarray}
where $Ei$ is the exponential integral function,
\begin{equation}
Ei(1, x) = \int_{1}^{\infty}du\ \frac{e^{-ux}}{x}\ \ ,\ x>0\ .
\end{equation}

Substituting (\ref{cont1}), (\ref{cont2}) and (\ref{cont3}) in (\ref{xcv3}) we have the interaction energy between a perfect conductor and a point-like steady charge,
\begin{eqnarray}
\label{energiaespcarga}
E_{PC}=-\frac{q^{2}}{2^{4}\pi}\frac{1}{R}\biggl(1+\Delta(mR)\biggr) \,
\end{eqnarray}
where we defined the function,
\begin{eqnarray}
\label{defDelta}
\Delta(mR)=mR\Bigg[\frac{K_{0}(2mR)}{mR}+\frac{K_{1}(2mR)}{(mR)^{2}}\cr
+\frac{\exp{(-2mR)}}{2(mR)^{3}}+\frac{\exp{(-2mR)}}{(mR)^{2}}-(mR)Ei(1,mR)\cr
+\frac{1}{2(mR)^{3}}+\frac{\pi}{2mR}[Y_{0}(2mR)-SH_{0}(2mR)]\cr
+\frac{\pi}{2(mR)^{2}}[SH_{1}(2mR)-Y_{1}(2mR)]\cr
+\exp{(-mR)}\Bigg(1-\frac{1}{mR}-\frac{2}{(mR)^{2}}-\frac{2}{(mR)^{3}}\Biggr)\Bigg] \ .
\end{eqnarray}

The result ({\ref{energiaespcarga}) is exact, but hard to be interpreted. The first term on the right hand side is the plate-charge interaction obtained in Maxwell electrodynamics, where the image method is valid, and does not involve the mass parameter $m$. The second term falls down when $mR$ increases. Once $m$ is a large quantity, for not so small distances $R$, this term is much smaller than the Coulombian one.

The interacting force between the plate and the charge is then given by,
\begin{eqnarray}
\label{FMC}
F_{PC}&=&-\frac{\partial E_{PC}}{\partial R}\cr\cr
&=&-\frac{q^{2}}{4\pi}\frac{1}{(2R)^{2}}[1+\Delta(mR)-mR\Delta'(mR)] \ ,
\end{eqnarray}
where the prime denotes derivative of $\Delta$ with respect to its argument. Notice that the force (\ref{FMC}) is the usual Coulombian interaction between the charge and its image, placed a distance $2R$ apart, and a correction term $m$-dependent. The term inside brackets on the right hand side of Eq. (\ref{FMC}) is always positive, so the expression (\ref{FMC}) is always negative, what means that the force is attractive.

Let us compare the force (\ref{FMC}) with the one obtained in Lee-Wick electrodynamics in $3+1$ dimensions for the interaction between two opposite charges, $q$ and $-q$, placed at a distance $2R$ apart,
\begin{equation}
\label{FCC}
F_{CC}=-\frac{q^{2}}{4\pi}\frac{1}{(2R)^{2}}[1-\exp(-2mR)-2mR\exp(-2mR)]\ .
\end{equation}
Eq. (\ref{FCC}) can be obtained with the results of reference \cite{BHN2013}. Specifically, from Eq. (16) of reference \cite{BHN2013} one can write the interaction energy between two opposite charges ($q$ and $-q$). So that, taking the gradient of this energy (with an overall minus sign) with respect to the distance of the charges and setting it equal to $2R$, we are taken to the result (\ref{FCC}). 

In the limit where $R\to0$, both forces (\ref{FMC}) and (\ref{FCC}) are finite, namely,
\begin{eqnarray}
\lim_{R\to0}F_{PC}=-\frac{q^{2}}{16\pi}\frac{3}{2}m^2\ \ ,\ \ 
\lim_{R\to0}F_{CC}=-\frac{q^{2}}{16\pi}2m^2\ .
\end{eqnarray}

It is interesting to notice that the image method is not valid for Lee-Wick electrodynamics for the conducting plate condition (\ref{condcondutor}). This fact can be seen from the interacting force between the plate and the charge (\ref{FMC}). The deviation from the image method behavior can be seen from the difference between (\ref{FMC}) and (\ref{FCC}) normalized by the Coulombian force, for convenience,
\begin{equation}
\delta(mR)=\frac{|F_{PC}|-|F_{CC}|}{[q^{2}/(4\pi)]\ [1/(2R)^{2}]}\ .
\end{equation}

\begin{figure}
\label{figura1}
 \centering
   \includegraphics[scale=0.40]{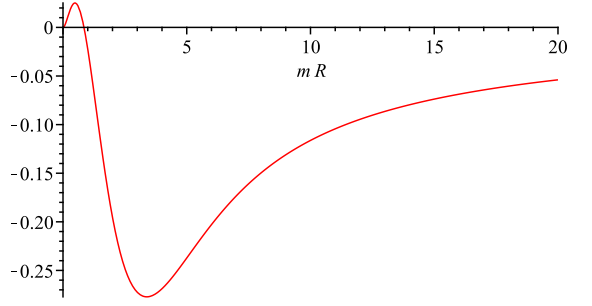}
   \caption{Plot for $\delta(mR)$.}
\end{figure}

The results of our numerical analysis suggest that the behavior of $\delta(mR)$ exhibits the same shape as the one observed in Fig.(1) up to $mR\to\infty$, with $\delta$ going to zero monotonically, with negative values, as $mR\to\infty$.

In the limit $mR\to\infty$ we have $\delta\to 0$. In the interval $0<mR\approx<0.85$ we have $\delta>0$ and the modulus of plate-charge interaction intensity is greater than the modulus of charge-image interaction. For $mR>\approx0.85$, $\delta<0$ and the charge-image interaction dominates, in modulus. In most situations the modulus of the force between the plate and the charge is lower than the modulus of the force between the charge and its image, as one can see from figure (1). It is also interesting to notice that the discrepancy between the plate-charge interaction force and the charge-image interaction exhibits a maximum around $mR\cong0,48$, a minimum around $mR\cong3.85$ and a zero around $mR\cong0.85$. 

Once $m$ must be a large quantity, only for very small values of the distance $R$ the force between the plate and the charge would have a stronger intensity than it would be if the image method were valid. So, it is relevant to consider if the maxima, minima and zero of the function $\delta$ can be really measured. Let us focus on the zero of $\delta$, which occurs at $R_{0}\cong0.85/m$. In recent works \cite{Accioly2010,TurcatiNeves,estimativas,estisospin} there are some estimates for lower bounds of Lee-Wick mass. The lowest estimates give, in order of magnitude, $m\sim 10GeV$. Working in $mks$ system, we can estimate in order of magnitude $R_{0}\cong 10^{-16}$, which is much smaller than the Compton wavelength $\lambda_{C}\cong10^{-12}$. In this scale many other effects might be taken into account, as the vacuum polarization, finite conductivity, rugosity and finite thickness of the plate and many others. So the behavior of the function $\delta$ next to the plate observed in Fig.(1) is not feasible to be reproduced in a laboratory.

In addition, for experimental setups, the distance scale relevant to investigate the interaction between classical objects are of micrometer order. So, taking the estimate of $m\sim 10GeV$, we have, $mR\cong10^{11}$ in $mks$, which is in a range where the function $\delta$ goes to zero with negative values. 

The estimate of $m\sim 10GeV$ is among the lowest ones. Other estimates \cite{estisospin} suggest a lower bound for $m$ in order of $\sim TeV$, what makes the phenomena related to the Lee-Wick Theory in the presence of a conducting plate even more negligible.

\section{The two field formalism}
\label{2campos}

We showed that when we consider the Lee-Wick field in the presence of a conducting surface, the gauge field must be submitted to the boundary conditions (\ref{condcondutor}) and the corresponding propagator (\ref{defDC}) can not be decomposed as the Maxwell propagator minus the Proca one, each one submitted to the conditions (\ref{condcondutor}) separately. As an immediate consequence, the image method is not valid for the Lee-Wick field.

In order to understand why this decomposition is no longer valid with the presence of a conducting plate, let us review how can we describe the Lee-Wick Lagrangian (\ref{defL}) without boundary conditions (without the presence of conductor) as the Maxwell Lagrangian minus the Proca one. Following a path integral approach, we can establish the Lee-Wick theory by a functional generator with an alternative Lagrangian of two coupled fields, $A^{\mu}$ and $S^{\mu}$, as follows
\begin{eqnarray}
\label{funcional2campos}
Z_{2}[J]=\int{\cal D}A{\cal D}S\exp\Biggl[i\int d^{4}x\ \frac{1}{2}\Biggl(\partial_{\mu}A^{\mu}\partial_{\nu}S^{\nu}\cr\cr
-\partial_{\gamma}A_{\mu}\partial^{\gamma}S^{\mu}-\frac{m^{2}}{4}A^{\mu}A_{\mu}-\frac{m^{2}}{4}S^{\mu}S_{\mu}\cr\cr
+\frac{m^{2}}{2}A^{\mu}S_{\mu}\Biggr)
-J^{\mu}A_{\mu}\Biggr]\ .
\end{eqnarray}

It is important to notice that the external source couples strictly to the $A^{\mu}$ field, and not to the $S^{\mu}$ field.

Integrating out the functional (\ref{funcional2campos}) on $S$ and preforming some mani\-pulations, we are taken to
\begin{eqnarray}
\label{dfg1}
Z_{2}[J]=Z_{S}[0]Z_{LW}[J]\ , 
\end{eqnarray}
where $Z_{LW}[J]$ is the standard functional generator for the Lee-Wick theory defined in (\ref{defZLW}) and 
\begin{eqnarray}
\label{defZS}
Z_{S}[0]=\int{\cal D}S\exp\Biggl(\frac{i}{2}\int d^{4}x -\frac{m^{2}}{4}S_{\mu}S^{\mu}\Biggr)
\end{eqnarray}
is the generating functional for $S^{\mu}$ without any external source. Once $Z_{S}[0]$ does not depend on the external source, it can be absorbed into a renormalization overall multiplicative constant and has no physical effect. So, from Eq. (\ref{dfg1}), we can write that
\begin{equation}
Z_{2}[J]=Z_{LW}[J]\ .
\end{equation}
It proves that without the presence of a conducting plate the two fields functional (\ref{funcional2campos}) is completely equivalent to (\ref{defZLW}), the functional generator obtained from the Lee-Wick Lagrangian (\ref{defL}) with a single field.

Now we perform, in the functional generator (\ref{funcional2campos}), the following change in the integrating field variables,
\begin{eqnarray}
\label{transf}
A^{\mu}=U^{\mu}+V^{\mu}\ \ ,\ \ S^{\mu}=U^{\mu}-V^{\mu}\cr\cr
U^{\mu}=\frac{1}{2}(A^{\mu}+S^{\mu})\ \ ,\ \ V^{\mu}=\frac{1}{2}(A^{\mu}-S^{\mu})\ .
\end{eqnarray}

The jacobian of the transformation (\ref{transf}) is a (divergent) constant which does not depend on the fields, thus it does not affect the functional generator because its contribution can be absorbed into an overall multiplicative constant. So ${\cal D}A{\cal D}S\cong{\cal D}U{\cal D}V$ and the functional (\ref{funcional2campos}), after some manipulations, reads
\begin{eqnarray}
\label{funcional2campossep}
Z_{2}[J]=\int{\cal D}U\exp\Biggl(i\int d^{4}x-\frac{1}{4}U_{\mu\nu}U^{\mu\nu}-J_{\mu}U^{\mu}\Biggr)\cr\cr
\int{\cal D}V\exp\Biggl(i\int d^{4}x\frac{1}{4}V_{\mu\nu}V^{\mu\nu}-\frac{m^{2}}{2}V^{\mu}V_{\mu}-J_{\mu}V^{\mu}\Biggr)\ ,\cr
\ 
\end{eqnarray}
where we defined
\begin{eqnarray}
U^{\mu\nu}=\partial^{\mu}U^{\nu}-\partial^{\nu}U^{\mu}\ \ \ ,\ \ \ 
V^{\mu\nu}=\partial^{\mu}V^{\nu}-\partial^{\nu}V^{\mu}\ .
\end{eqnarray}

Notice that (\ref{funcional2campossep}) can be decomposed as the product of a Maxwell functional generator for the $U$ field with a Proca functional generator with an overall minus sign in the Lagrangian for the $V$ field. The external source $J^{\mu}$ couples to both fields, $U$ and $V$. It proves that (\ref{funcional2campos}), and so (\ref{defZLW}), is completely equivalent to a Maxwell functional generator minus a Proca functional generator with an overall minus sign in the Lagrangian.

Now, let us consider the two field theory (\ref{funcional2campos}) with the presence of a conducting plate. The Lorentz force must vanish on the plate, as discussed in section (\ref{secpropagador}). Once only the $A^{\mu}$ field couples to the external source $J^{\mu}$, the conditions which vanishes the Lorentz force on the plate is given by Eq. (\ref{condcondutor}), that is imposed only on the $A^{\mu}$ field. In this case we have to restrict the functional integral over field configurations which satisfy (\ref{condcondutor}). This constraint over the field configuration can be attained by inserting the delta functional (\ref{deltafuncional}) in the integral (\ref{funcional2campos}), in the same manner we have done in section (\ref{secpropagador}),
\begin{eqnarray}
\label{funcional2camposCC}
Z_{2,C}[J]=\int{\cal D}A{\cal D}S\ \delta[F^{*}_{\ 3\nu}(x)|_{x^{3}=a}]\cr\cr
\exp\Biggl[i\int d^{4}x\ \frac{1}{2}\Biggl(\partial_{\mu}A^{\mu}\partial_{\nu}S^{\nu}-\partial_{\gamma}A_{\mu}\partial^{\gamma}S^{\mu}\cr\cr
-\frac{m^{2}}{4}A^{\mu}A_{\mu}-\frac{m^{2}}{4}S^{\mu}S_{\mu}+\frac{m^{2}}{2}A^{\mu}S_{\mu}\Biggr)
-J^{\mu}A_{\mu}\Biggr]\ ,
\end{eqnarray}
where $Z_{2,C}[J]$ stands for the two field functional generator with the presence of a conducting plate.

It is important to point out that in Eq. (\ref{funcional2camposCC}) the integral is performed over all $A$-field configurations. The constraint (\ref{condcondutor}) is attained by the delta functional $\delta[F^{*}_{\ 3\nu}(x)|_{x^{3}=a}]$. There is no constraint on the $S$-field.
 
Integrating out the functional (\ref{funcional2camposCC}) on the $S$ variable, and performing some manipulations, we are taken to
\begin{eqnarray}
\label{dfg2}
Z_{2,C}[J]=Z_{S}[0]Z_{C}[J]\ ,
\end{eqnarray}
where $Z_{C}[J]$, defined in (\ref{Zfinal}), is the Lee-Wick functional generator with boundary conditions (\ref{condcondutor}) and $Z_{S}[0]$ is the functional generator (\ref{defZS}) for the field $S$ with no boundary conditions and no external source. As discussed previously, in this case $Z_{S}[0]$ has no physical effect and we can write from Eq. (\ref{dfg2})
\begin{equation}
Z_{2,C}[J]=Z_{C}[J]\ .
\end{equation}
It proves that the boundary conditions can be imposed in both approaches, namely, the one with two fields the other with one single field. The physical results are the same in both cases.

Now we perform the change of integrating variables (\ref{transf}) in the two field functional with a conducting plate (\ref{funcional2camposCC}), in order to know how the boundary conditions (\ref{condcondutor}) are implemented in terms of the fields $U$ and $V$. For this task we substitute Eq. (\ref{deltafuncional2}) in (\ref{funcional2camposCC}), similarly to what we have done in section (\ref{secpropagador}), and use the definitions (\ref{transf}), what leads to
\begin{eqnarray}
\label{dfg3}
Z_{2,C}[J]=\int{\cal D}U{\cal D}V{\cal D}B\cr\cr
\exp\Biggl(-i\int d^{4}x\delta(x^{3}-a)\epsilon_{3}^{\ \nu\alpha\beta}U_{\beta}(x)\partial_{\alpha}B_{\nu}(x_{\|})\Biggr)\cr\cr
\exp\Biggl(-i\int d^{4}x\delta(x^{3}-a)\epsilon_{3}^{\ \nu\alpha\beta}V_{\beta}(x)\partial_{\alpha}B_{\nu}(x_{\|})\Biggr)\cr\cr
\exp\Biggl(\frac{i}{2\gamma}\int d^{3}x_{\parallel}d^{3}y_{\parallel}B^{\mu}(x_{\parallel})
\frac{\partial^{2}Q(x,y)}{\partial x_{\parallel}^{\mu}\partial y_{\parallel}^{\nu}}\Bigg|_{\substack{x^{3}=a\\y^{3}=a}}B^{\nu}(y_{\parallel})\Biggr)\cr\cr
\exp\Biggl(i\int d^{4}x\frac{1}{4}V_{\mu\nu}V^{\mu\nu}-\frac{m^{2}}{2}V^{\mu}V_{\mu}-J_{\mu}V^{\mu}\Biggr)\cr\cr
\exp\Biggl(i\int d^{4}x-\frac{1}{4}U_{\mu\nu}U^{\mu\nu}-J_{\mu}U^{\mu}\Biggr) ,
\end{eqnarray}
where $Q(x,y)$ is an arbitrary function we will choose conveniently and $\gamma$ is a gauge parameter.

From now on, we shall work in the gauge where $\gamma=1$.

Before we start to solve the functional integral (\ref{dfg3}), it is important to notice that it exhibits just one single auxiliary field $B$ coupled to both fields, $U$ and $V$. 

The first and second exponentials in (\ref{dfg3}) couples, respectively, the fields $U$ and $V$ to the auxiliary field $B^{\mu}$. 
We can decouple these fields in the same way we have done in section (\ref{secpropagador}), with the translations
\begin{eqnarray}
U_{\beta}(x)\to U_{\beta}(x)\cr
+\int d^{4}y D^{(M)}_{\beta\alpha}(x,y)\delta(y^{3}-a)\epsilon_{3}^{\ \nu\gamma\alpha}\partial_{\gamma}B_{\nu}(y_{\|})\cr\cr  
V_{\beta}(x)\to V_{\beta}(x)\cr
-\int d^{4}y D^{(P)}_{\beta\alpha}(x,y)\delta(y^{3}-a)\epsilon_{3}^{\ \nu\gamma\alpha}\partial_{\gamma}B_{\nu}(y_{\|})\ ,
\end{eqnarray}
where we defined the Maxwell and Proca propagators, respectively, as follows
\begin{eqnarray}
\label{defDmDp}
D^{(M)}_{\beta\alpha}(x,y)&=&\int\frac{d^{4}p}{(2\pi)^{4}}-\frac{1}{p^{2}}\eta^{\mu\nu}e^{-ip(x-y)}\cr\cr
D^{(P)}_{\beta\alpha}(x,y)&=&\int\frac{d^{4}p}{(2\pi)^{4}}\frac{1}{m^{2}-p^{2}}\Biggl(\eta^{\mu\nu}-\frac{p^{\mu}p^{\nu}}{m^{2}}\Biggr)e^{-ip(x-y)} .\cr
&\ & 
\end{eqnarray}

With these considerations we can rewrite the functional generator (\ref{dfg3}), after some manipulations, in the form
\begin{eqnarray}
\label{dfg5}
Z_{2,C}[J]={\bar Z}_{2,C}[J]\times\cr\cr
\int{\cal D}U\exp\Biggl(i\int d^{4}x-\frac{1}{4}U_{\mu\nu}U^{\mu\nu}-J_{\mu}U^{\mu}\Biggr)\cr\cr
\int{\cal D}V\exp\Biggl(i\int d^{4}x\frac{1}{4}V_{\mu\nu}V^{\mu\nu}-\frac{m^{2}}{2}V^{\mu}V_{\mu}-J_{\mu}V^{\mu}\Biggr)\cr\cr
\end{eqnarray}
where we defined
\begin{eqnarray}
\label{defZ2Cbar}
{\bar Z}_{2,C}[J]=\int{\cal D}B\exp\Bigg[\frac{i}{2}\int d^{4}xd^{4}y\Bigg(\frac{\partial^{2}Q(x,y)}{\partial x_{\|\rho}\partial y_{\|\beta}}\cr\cr
-\epsilon_{3}^{\ \rho\tau\mu}\epsilon_{3}^{\ \beta\gamma\alpha}\frac{\partial^{2}[D^{(M)}_{\mu\alpha}(x,y)-D^{(P)}_{\mu\alpha}(x,y)]}{\partial x_{\|}^{\tau}\partial y_{\|}^{\gamma}}\Bigg)\cr\cr
\times\delta(x^{3}-a)\delta(y^{3}-a)B_{\rho}(x_{\|})B_{\beta}(y_{\|})\Bigg]\cr\cr
\exp\Bigg[i\int d^{4}x\delta(x^{3}-a)B_{\nu}(x_{\|})\cr
\times\Bigg(\int d^{4}y\epsilon_{3}^{\ \nu\gamma\alpha}J_{\mu}(x)\frac{\partial}{\partial x^{\gamma}}\Big(D^{(M)\mu\nu}(x,y)\cr\cr
-D^{(P)\mu\nu}(x,y)\Big)\Bigg)\Bigg]\cr
\ 
\end{eqnarray}

Choosing the function $Q(x,y)$ as the one defined in Eq. (\ref{defQ}), using definitions (\ref{defDmDp}) and solving the functional integral in Eq. (\ref{defZ2Cbar}), which is quadratic in $B^{\mu}$, it can be shown that ${\bar Z}_{2,C}[J]$ gives the same result found in Eq. (\ref{barZcalculado}) for $\bar Z$ (up to an overall multiplicative constant which does not depend on $J$), it is,
\begin{eqnarray}
\label{dfg6}
{\bar Z}_{2,C}[J]&\sim&\bar Z[J]\cr\cr
&=&{\bar Z}_{2,C}[0]e^{-\frac{i}{2}\int d^{4}x d^{4}y J^{\mu}(x){\bar D}_{\mu\nu}(x,y)J^{\nu}(y)}\ .
\end{eqnarray}

Substituting (\ref{dfg6}) in (\ref{dfg5}) and integrating out on $U$ and $V$ it can be shown that the two field functional generator with the presence of conducting plate (\ref{dfg5}) is equal to the right hand side of (\ref{Zfinal}), namely
\begin{eqnarray}
Z_{2,C}[J]&=&Z_{2,C}[0]\exp\Bigg[-\frac{i}{2}\int d^{4}x d^{4}y J^{\mu}(x)\Bigl(D_{\mu\nu}(x,y)\cr
&\ &+{\bar D}_{\mu\nu}(x,y)\Bigr)J^{\nu}(y)\Bigg]\ .
\end{eqnarray}

It proves again that, with the boundary conditions (\ref{condcondutor}), the two field formalism leads to the same results as the ones obtained from the formalism with just one single field.

\subsection*{The wrong boundary condition}
\label{errado}

In spite of not being the true boundary condition imposed on the fields by the presence of a conducting surface, we could ask what kind of theory we would have by imposing, to each field $A^{\mu}$ and $S^{\mu}$, a boundary condition similar to (\ref{condcondutor}). Before we answer this question, we point out, once more, that the matter source $J$ couples just with the $A$ field, so is this field which produces the Lorentz force on the charged particles. On a conducting surface, the Lorentz force must vanish, what is attained by imposing boundary conditions only on the $A$-field and not on the $S$-field. 

In order to impose the conditions (\ref{condcondutor}) to the fields $A^{\mu}$ and $S^{\mu}$, we must insert two delta functionals inside the integral (\ref{funcional2campos}), as follows
\begin{eqnarray}
\label{lkj1}
Z_{2,NC}[J]=\int{\cal D}A{\cal D}S\delta[F^{*}_{\ 3\nu}(x)|_{x^{3}=a}]\delta[S^{*}_{\ 3\nu}(x)|_{x^{3}=a}]\cr\cr
\exp\Biggl[i\int d^{4}x\ \frac{1}{2}\Biggl(\partial_{\mu}A^{\mu}\partial_{\nu}S^{\nu}-\partial_{\gamma}A_{\mu}\partial^{\gamma}S^{\mu}\cr\cr
+\frac{m^{2}}{2}A^{\mu}S_{\mu}-\frac{m^{2}}{4}A^{\mu}A_{\mu}-\frac{m^{2}}{4}S^{\mu}S_{\mu}\Biggr)-J^{\mu}A_{\mu}\Biggr] ,
\end{eqnarray}
where $S_{\mu\nu}^{*}=\epsilon_{\mu\nu\alpha\beta}\partial^{\alpha}S^{\beta}$ and the sub-index $NC$ stands for boundary conditions which does not represent a conducting surface. 

Each delta functional in (\ref{lkj1}) have an integral representation like the one in Eq. (\ref{deltafuncional2}), namely,
\begin{eqnarray}
\label{2deltasfuncionais}
\delta[{F^{*}}_{3\nu}(x)|_{x^{3}=a}]=\cr\cr
N\int{\cal D}B\exp\Biggl(-i\int d^{4}x\delta(x^{3}-a)A_{\beta}(x)\epsilon_{3}^{\ \nu\alpha\beta}\partial_{\alpha}B_{\nu}(x_{\parallel})\Biggr)\cr\cr
\exp\Biggl(\frac{i}{2\gamma_{A}}\int d^{3}_{\parallel}xd^{3}y_{\parallel} B^{\mu}(x_{\parallel})
\frac{\partial^{2}Q_{A}(x,y)}{\partial x_{\parallel}^{\mu}\partial y_{\parallel}^{\nu}}\Bigg|_{\substack{x^{3}=a\\y^{3}=a}}B^{\nu}(y_{\parallel})\Biggr),\cr\cr
\delta[{S^{*}}_{3\nu}(x)|_{x^{3}=a}]=\cr\cr
N\int{\cal D}C\exp\Biggl(-i\int d^{4}x\delta(x^{3}-a)S_{\beta}(x)\epsilon_{3}^{\ \nu\alpha\beta}\partial_{\alpha}C_{\nu}(x_{\parallel})\Biggr)\cr\cr
\exp\Biggl(\frac{i}{2\gamma_{S}}\int d^{4}x_{\parallel}d^{4}y_{\parallel} C^{\mu}(x_{\parallel})
\frac{\partial^{2}Q_{S}(x,y)}{\partial x_{\parallel}^{\mu}\partial y_{\parallel}^{\nu}}\Bigg|_{\substack{x^{3}=a\\y^{3}=a}}C^{\nu}(y_{\parallel})\Biggr),\cr
\ 
\end{eqnarray}
where $\gamma_{A}$ and $\gamma_{S}$ are gauge parameters and $Q_{A}(x,y)$ and $Q_{S}(x,y)$ are arbitrary functions we shall choose conveniently.

At this point one comment is in order. When we insert the integrals (\ref{2deltasfuncionais}) into (\ref{lkj1}) we shall have two auxiliary fields, $B$ and $C$. In Eq. (\ref{dfg3}) we have just one auxiliary field.

Using the gauges where $\gamma_{A}=\gamma_{S}=1/2$, taking the functions $Q_{A}(x,y)=Q_{S}(x,y)$ equal to the one of Eq. (\ref{defQ}), substituting Eq's (\ref{2deltasfuncionais}) in (\ref{lkj1}) and performing the change of field variables (\ref{transf}) and
\begin{eqnarray}
B_{+}=B+C\ \ \ ,\ \ \ B_{-}=B-C\cr\cr
B=\frac{1}{2}(B_{+}+B_{-})\ \ \ ,\ \ \ C=\frac{1}{2}(B_{+}-B_{-})\ ,
\end{eqnarray}
whose jacobian does not depend on any field, we have
\begin{equation}
\label{nhy1}
Z_{2,NC}[J]=Z_{M,C}[J]Z_{-P,C}[J]
\end{equation}
where
\begin{eqnarray}
\label{defZMC}
Z_{M,C}[J]=\int{\cal D}U\exp\Biggl(i\int d^{4}x-\frac{1}{4}U_{\mu\nu}U^{\mu\nu}-J_{\mu}U^{\mu}\Biggr)\cr\cr
\int{\cal D}B_{+}\exp\Biggl(-i\int d^{4}x\delta(x^{3}-a)U_{\beta}(x)\epsilon_{3}^{\ \nu\alpha\beta}\partial_{\alpha}B_{+\nu}(x_{\parallel})\Biggr)\cr\cr
\exp\Biggl(\frac{i}{2}\int d^{3}_{\parallel}xd^{3}y_{\parallel} B^{\mu}_{+}(x_{\parallel})
\frac{\partial^{2}Q(x,y)}{\partial x_{\parallel}^{\mu}\partial y_{\parallel}^{\nu}}\Bigg|_{\substack{x^{3}=a\\y^{3}=a}}B^{\nu}_{+}(y_{\parallel})\Biggr)\cr
\ 
\end{eqnarray}
is the functional generator for the Maxwell field $U$ submitted to the boundary condition $U^{*}_{3\mu}=0$ on the conducting surface $x^{3}=0$ and
\begin{eqnarray}
\label{defZ-PMC}
Z_{-P,C}[J]=\cr\cr
\int{\cal D}V\exp\Biggl(i\int d^{4}x\frac{1}{4}V_{\mu\nu}V^{\mu\nu}-\frac{m^{2}}{2}V^{\mu}V_{\mu}-J_{\mu}V^{\mu}\Biggr)\cr\cr
\int{\cal D}B_{-}\exp\Biggl(-i\int d^{4}x\delta(x^{3}-a)V_{\beta}(x)\epsilon_{3}^{\ \nu\alpha\beta}\partial_{\alpha}B_{-\nu}(x_{\parallel})\Biggr)\cr\cr
\exp\Biggl(\frac{i}{2}\int d^{3}_{\parallel}xd^{3}y_{\parallel} B^{\mu}_{-}(x_{\parallel})
\frac{\partial^{2}Q(x,y)}{\partial x_{\parallel}^{\mu}\partial y_{\parallel}^{\nu}}\Bigg|_{\substack{x^{3}=a\\y^{3}=a}}B^{\nu}_{-}(y_{\parallel})\Biggr)\cr
\ 
\end{eqnarray}
is the functional generator for the Proca field $V$, with an overall minus sign in the Lagrangian, submitted to the boundary condition $V^{*}_{3\mu}=0$ on the conducting surface $x^{3}=0$.

Both integrals (\ref{defZMC}) and (\ref{defZ-PMC}) can be calculated following the same procedure employed in the previous sections. Working in the Lorentz gauge for the Maxwell field, the results are 
\begin{eqnarray}
\label{ZMC-PC}
Z_{M,C}[J]=\exp\Biggl(\frac{i}{2}\int d^{4}xd^{4}yJ_{\mu}(x)\Bigl(D_{M}^{\mu\nu}(x,y)\cr\cr
+{\bar D}_{M}^{\mu\nu}(x,y)\Bigl)J_{\nu}(y)\Biggr)\cr\cr
Z_{-P,C}[J]=\exp\Biggl(\frac{i}{2}\int d^{4}xd^{4}yJ_{\mu}(x)\Bigl(-D_{P}^{\mu\nu}(x,y)\cr\cr
-{\bar D}_{P}^{\mu\nu}(x,y)\Bigl)J_{\nu}(y)\Biggr)
\end{eqnarray}
where $D_{P}^{\mu\nu}(x,y)$ and $D_{M}^{\mu\nu}(x,y)$ stand for the free (without boundary conditions) Proca and Maxwell propagators, respectively,
\begin{eqnarray}
D_{P}^{\mu\nu}(x,y)&=&\int\frac{d^{4}p}{(2\pi)^{4}}\frac{-1}{p^{2}-m^{2}}\Biggl(\eta^{\mu\nu}-\frac{p^{\mu}p^{\nu}}{m^{2}}\Biggr)e^{-ip(x-y)}\cr\cr
D_{M}^{\mu\nu}(x,y)&=&\int\frac{d^{4}p}{(2\pi)^{4}}\frac{-1}{p^{2}}\eta^{\mu\nu}e^{-ip(x-y)}
\end{eqnarray}
and we defined the functions
\begin{eqnarray}
{\bar D}_{P}^{\mu\nu}(x,y)=\int\frac{d^{3}p_{\|}}{(2\pi)^{3}}-\frac{i}{2\Gamma}\Bigg(\eta^{\mu\nu}_{\|}-\frac{p_{\|}^{\mu}p_{\|}^{\mu}}{p_{\|}^{2}}\Bigg)\cr\cr
\times e^{i\Gamma|x^{3}-a|}e^{i\Gamma|y^{3}-a|}e^{-ip_{\|}(x_{\|}-y_{\|})}\cr\cr
{\bar D}_{M}^{\mu\nu}(x,y)=\int\frac{d^{3}p_{\|}}{(2\pi)^{3}}-\frac{i}{2L}\Bigg(\eta^{\mu\nu}_{\|}-\frac{p_{\|}^{\mu}p_{\|}^{\mu}}{p_{\|}^{2}}\Bigg)\cr\cr
\times e^{iL|x^{3}-a|}e^{iL|y^{3}-a|}e^{-ip_{\|}(x_{\|}-y_{\|})}
\end{eqnarray}
which give the corrections for the above propagators due to the boundary conditions $V^{*}_{3\mu}=0$ and $U^{*}_{3\mu}=0$, respectively.

Substituting (\ref{ZMC-PC}) in (\ref{nhy1}) we finally have
\begin{eqnarray}
\label{funcionalerrado}
Z_{2,NC}[J]=\cr
\exp\Bigg[-\frac{i}{2}\int d^{4}xd^{4}y\ J_{\mu}(x)\Big[\Big(D_{M}^{\mu\nu}(x,y)+{\bar D}_{M}^{\mu\nu}(x,y)\Big)\cr\cr
-\Big(D_{P}^{\mu\nu}(x,y)+{\bar D}_{P}^{\mu\nu}(x,y)\Big)\Big]J_{\nu}(y)\Bigg] .\ 
\end{eqnarray}

From (\ref{funcionalerrado}) one can identify the propagator of the model as composed by the Maxwell propagator in the presence of a conducting plane minus a Proca propagator (with an overall minus sign) submitted to the condition $V^{*}_{3\mu}=0$ on the conducting plane. In this case, it is not difficult to show that the image method is valid and all physical phenomena of the theory, with the wrong boundary conditions, can be decomposed as the corresponding ones obtained in the Maxwell Theory in the presence of a conducting plate minus the ones obtained in the Proca Theory, with the condition $V^{*}_{3\mu}=0$ on the plate.

\section{Conclusions and final remarks}
\label{conclusoes}

In this paper we developed the Lee-Wick electrodynamics in the presence of a single perfectly conducting plate. The boundary condition imposed on the gauge field is the one which vanishes, on the plate, the dual field strength normal to the plate. That is justified because it is the condition which vanishes the Lorentz force on a given electric charge placed on the plate, what is in accordance with the notion of a perfect conductor. Using functional methods of Quantum Field Theory, we calculated the gauge field propagator and showed that it is not the composition of the Maxwell propagator minus the Proca one, both with the presence of a conducting plate. This fact makes the physical phenomena in the Lee-Wick electrodynamics not usual even classically.

With the obtained propagator any quantity, quantum or classical, related to the gauge field in Lee-Wick electrodynamics could be computed. We calculated the interaction energy between a stationary point charge and a conducting plate. The counterpart of this result in Maxwell or Proca electrodynamics is the plate-charge interaction, which leads to the well known image method. From our results it is shown that in the Lee-Wick electrodynamics the image method is not valid anymore.

We made a discussion regarding the two field formalism for the Lee-Wick electrodynamics, with the presence of a conducting plate, and showed that the physical results are the same as the ones obtained from the one field formalism. We also discussed, in the two field formalism, what would be the boundary conditions which make the theory equivalent to the Maxwell electrodynamics with the presence of the conducting plate minus the Proca Theory with boundary conditions. We showed that this boundary conditions are not the true ones imposed by a conducting surface.  

The Lee-Wick electrodynamics with the presence of two conducting parallel plates can be developed with the same methods employed in this paper. It is another interesting situation where we can study true quantum phenomena, as the Casimir effect, for instance \cite{BNH}.

\

{\bf Acknowledgments}

\noindent

F.A Barone and A.A. Nogueira are very grateful to CNPq and Capes (Brazilian agencies) for financial support. The authors thank to F.E. Barone for reviewing the paper and for very pertinent comments.

\appendix

\section{Delta functional}
\label{calculodelta}

In this appendix we obtain expression (\ref{deltafuncional2}). First, due to the gauge invariance (\ref{calibreB}), we have to extract the infinite gauge volume from the integral (\ref{deltafuncional}). For this task we follow the Faddeev-Popov trick, fixing the covariant gauge
\begin{equation}
F[B^{\mu}(x_{\|})]=\partial^{\nu}_{\|}B_{\nu}(x_{\|})=f(x_{\|})\ ,
\end{equation}
where $f(x_{\|})$ is an arbitrary function. The corresponding Faddeev-Popov determinant does not depend on the field $B^{(k)}_{\nu}(x_{\|})$ and, so, does not contribute to the integral. In this way, we can write (\ref{deltafuncional}) in the form
\begin{eqnarray}
\label{zxc1}
\delta[n_{\mu}{F^*}^{\mu\nu}(x)|_{a}]\sim\int{\cal D}{B}\delta\Bigl[F[B_{\nu}(x_{\|})]-f(x_{\|})\Bigr]\cr\cr
\exp\Biggl[i\int d^{4}x \delta(x^{3}-a)B_{\nu}(x){F^*}_{3}^{\ \nu}(x)\Biggr]\ .
\end{eqnarray}

Now we integrate by parts the argument of the exponential and use t'Hooft trick, multiplying both sides of (\ref{zxc1}) by a convergent functional of $f(x_{\|})$ and integrating over $f(x_{\|})$, as follows
\begin{eqnarray}
\label{apend1}
\delta[n_{\mu}{F^*}^{\mu\nu}(x)|_{a}]=N\int{\cal D}f\int{\cal D}{B}\ \delta\Bigl[F[B_{\nu}(x_{\|})]-f(x_{\|})\Bigr]\cr\cr
\exp\Biggl[-i\int d^{4}x \delta(x^{3}-a)A_{\beta}(x)n_{\mu}\epsilon^{\mu\nu\alpha\beta}\partial_{\alpha}B_{\nu}(x)\Biggr]\cr\cr
\exp\Bigg[\frac{i}{2\gamma}\int d^{4}xd^{4}y\delta(x^{3}\!\!-a)f(x_{\|})Q(x,y)f(y_{\|})\delta(y^{3}\!\!-a)\Bigg],\cr
\end{eqnarray}
where $Q(x,y)$ is an arbitrary function, $N$ is a constant and $\gamma$ is an arbitrary gauge constant.

Performing the functional integral in $f$ and two integration by parts in Eq. (\ref{apend1}) we are taken to Eq. (\ref{deltafuncional2}).

\section{Integrals (\ref{asd4})}
\label{integrais}

In this appendix we compute the integrals (\ref{asd4}). For this task, it is enough to find out just the first Eq. (\ref{asd4}) because the second one is a special case of the first one, with $m=0$. 

Using the fact that there is a small negative imaginary part for the momentum square, as discussed just below Eq. (\ref{propagador2}), we have
\begin{eqnarray}
\int\frac{dp^{3}}{2\pi}\frac{1}{p^{2}-m^{2}}e^{ip^{3}x^{3}}=\lim_{\varepsilon\to0}\int\frac{dp^{3}}{2\pi}\frac{1}{p_{\|}^{2}-(p^{3})^{2}+i\varepsilon}e^{ip^{3}x^{3}}\cr\cr
\cong-\lim_{\varepsilon\to0}\int\frac{dp^{3}}{2\pi}\frac{1}{(p^{3})^{2}-L^{2}-i\varepsilon}e^{ip^{3}x^{3}}.\cr
\ 
\end{eqnarray}
Now, we solve the integral in the second line of the above equation by using residue theorem and take the limit $\varepsilon\to0$. The result is the first Eq. (\ref{asd4}).

\section{The limit $m\to\infty$}
\label{limitepropagador}

In this appendix we calculate the limit $m\to\infty$ of the propagator (\ref{defDC}). Once the parameter $m$ is present only in ${\bar D}_{\mu\nu}(x,y)$, it is enough to consider only this term. First we note that the correction to the propagator (\ref{defbarD}) can be written as
\begin{equation}
\label{defbarD2}
{\bar D}^{\mu\nu}(x,y)={\cal O}^{\mu\nu}{\cal I}(x,y)\ ,
\end{equation}
where we defined the differential operator ${\cal O}=\eta_{\|}^{\mu\nu}\partial_{\lambda\|}\partial^{\lambda}_{\|}-\partial^{\mu}_{\|}\partial^{\nu}_{\|}$ and the integral 
\begin{eqnarray}
\label{defcalI}
{\cal I}=\int\frac{d^{3}p_{\|}}{(2\pi)^{3}}\frac{i}{2}\frac{1}{p^{2}_{\|}}\frac{1}{\frac{1}{L}-\frac{1}{\Gamma}}e^{-ip_{\|}(x_{\|}-y_{\|})}\cr\cr
\Biggl(\frac{e^{iL|x^{3}-a|}}{L}-\frac{e^{i\Gamma|x^{3}-a|}}{\Gamma}\Biggr)
\Biggl(\frac{e^{iL|y^{3}-a|}}{L}-\frac{e^{i\Gamma|y^{3}-a|}}{\Gamma}\Biggr) .
\end{eqnarray}

Now we make a Wick rotation only in the parallel coordinates, in the integral (\ref{defcalI}), with $k^{4}=-ip^{0}$, $k^{1}=p^{1}$,\break $k^{2}=p^{2}$, $X^{4}=ix^{0}$, $X^{1}=x^{1}$ and $X^{2}=x^{2}$, and\break define the Euclidean 3-vectors ${\bf k}=(k^{1},k^{2},k^{4})$,\break ${\bf X}=(X^{1},X^{2},X^{4})$ (and similarly for $y^{\mu}_{\|}$), what leads to
\begin{eqnarray}
{\cal I}=\int\frac{d^{3}{\bf k}}{(2\pi)^{3}}\frac{e^{-ik^{4}(X^{4}-Y^{4})}}{2{\bf k}^{2}}
\frac{e^{ik^{1}(X^{1}-Y^{1})}e^{ik^{2}(X^{2}-Y^{2})}}{\frac{1}{i\sqrt{{\bf k}^{2}}}-\frac{1}{i\sqrt{{\bf k}^{2}+m^{2}}}}\cr\cr
\Biggl(\frac{1}{i\sqrt{{\bf k}^{2}}}e^{-\sqrt{{\bf k}^{2}}|x^{3}-a|}-\frac{1}{i\sqrt{{\bf k}^{2}+m^{2}}}e^{-\sqrt{{\bf k}^{2}+m^{2}}|x^{3}-a|}\Biggr)\cr\cr
\Biggl(\frac{1}{i\sqrt{{\bf k}^{2}}}e^{-\sqrt{{\bf k}^{2}}|y^{3}-a|}-\frac{1}{i\sqrt{{\bf k}^{2}+m^{2}}}e^{-\sqrt{{\bf k}^{2}+m^{2}}|y^{3}-a|}\Biggr) .
\end{eqnarray}

Taking the limit $m\to\infty$ in the above expression,
\begin{eqnarray}
\label{zxc2}
\lim_{m\to\infty}{\cal I}=\int\frac{d^{3}{\bf k}}{(2\pi)^{3}}\frac{1}{2{\bf k}^{2}}e^{-ik^{4}(X^{4}-Y^{4})}\cr\cr
e^{ik^{1}(X^{1}-Y^{1})}e^{ik^{2}(X^{2}-Y^{2})}\frac{e^{-\sqrt{{\bf k}^{2}}|x^{3}-a|}e^{-\sqrt{{\bf k}^{2}}|y^{3}-a|}}{i\sqrt{{\bf k}^2}}\ .
\end{eqnarray}

Performing an inverse Wick rotation back to Minkowski space on Eq. (\ref{zxc2}), substituting the result in to Eq. (\ref{defbarD2}) and acting with the operator ${\cal O}^{\mu\nu}$, we obtain
\begin{eqnarray}
\lim_{m\to\infty}{\bar D}^{\mu\nu}(x,y)=\int\frac{d^{3}p_{\|}}{(2\pi)^{3}}-\frac{i}{2}\Biggl(\eta^{\mu\nu}_{\|}-\frac{p^{\mu}_{\|}p^{\nu}_{\|}}{p_{\|}^{2}}\Biggr)\cr\cr
e^{-ip_{\|}(x_{\|}-y_{\|})}\frac{e^{iL(|x^{3}-a|+|y^{3}-a|)}}{L}\ ,
\end{eqnarray}
which is the propagator for the Maxwell field with the presence of a perfectly conducting plate placed at position $x^{3}=a$ \cite{Bordag}.



\end{document}